\documentclass[12pt]{article}
\usepackage{esfconf}
\usepackage{latexsym}

\begin{document}
\title{Low-Energy Effective Action\\ in N=4 Super Yang-Mills\\
Theories}

\authors{I.L. Buchbinder\adref{1,2}}

\addresses{\1ad
Instituto de F\'{i}sica, Universidade de S\~{a}o Paulo\\
P.O. Box 66318, 05315-970, S\~{a}o Paulo, Brasil,
\nextaddress
\2ad
Department of
Theoretical Physics Tomsk State Pedagogical University \\
Tomsk, 634041, Russia}

\maketitle

\begin{abstract}
We review a recent progress in constructing
low-energy effective action in N=4 super Yang-Mills theories. Using
harmonic superspace approach we consider N=4 SYM in terms of
unconstrained N=2 superfield and apply N=2 background field method to
finding effective action for N=4 SU(n) SYM broken down to U(n)$^{n-1}$.
General structure of leading low-energy corrections to effective action is
discussed.
\end{abstract}

\section{Introduction}

Low-energy structure of quantum supersymmetric field theories is
described by the effective lagrangians of two types: chiral and general
or holomorphic and non-holomorphic.  Non-holomorphic or general
contributions to effective action are given by integrals over full
superspace while holomorphic or chiral contributions are given by
integrals over chiral subspace of superspace. As a result the effective
action in low-energy limit is defined by the chiral superfield ${\cal
F}$ which is called holomorphic or chiral effective potential and real
superfield ${\cal H}$ which is called non-holomorphic or general
effective potential. The specific feature of these potentials is that
they do not depend on covariant derivatives of the superfields.

We point out that a possibility of holomorphic corrections to effective
action was firstly demonstrated in   [1] ( see also [2]) for N=1 SUSY
and in [3] for N=2 SUSY.  The modern interest to structure of
low-energy effective action in extended supersymmetric theories was
inspired by the seminal papers [4] where exact instanton contribution
to holomorphic effective potential has been found for N=2 SU(2) super
Yang-Mills theory.  These results have later been extended for various
gauge groups and for coupling to matter (see f.e. [5]).  One can show
that in generic N=2 SUSY models namely the holomorphic effective
potential is leading low-energy contribution. Non-holomorphic potential
is next to leading correction. A detailed investigation of structure of
low-energy effective action for various N=2 SUSY theories has been
undertaken in [6-9].

A further study of quantum aspects of
supersymmetric field models leads to problem of effective action in N=4
SUSY theories. These theories being maximally extended global
supersymmetric models possess the remarkable properties on quantum
level:

\begin{itemize}
\item[(i)] N=4 super Yang-Mills model is finite quantum field theory,

\item[(ii)] N=4 super Yang-Mills model is superconformal invariant
theory and hence, its effective action can not depend on any scale.
These properties allow to analyze a general form of low-energy
effective action and see that it changes drastically in compare with
generic N=2 super Yang-Mills theories.
\end{itemize}

Analysis of structure of low-energy effective action in N=4 SU(2) SYM
model spontaneously broken down to U(1) has been fulfilled in recent
paper by Dine and Seiberg [10]. They have investigated a part of
effective action depending on N=2 superfield strengths $W$, $\bar{W}$
and shown
\begin{itemize}
\item[(i)]
Holomorphic quantum corrections are trivial in N=4 SYM. Therefore,
namely non-holomorphic effective potential is leading low-energy
contribution to effective action,
\item[(ii)] Non-holomorphic
effective potential ${\cal H}(W,\bar W)$ can be found on the base of
the properties of quantum N=4 SYM theory up to a coefficient.
All perturbative or non-perturbative corrections do not influence on
functional form of ${\cal H}(W,\bar W)$ and concern only this
coefficient.
\end{itemize}

The approaches to direct calculation of non-holomorphic effective
potential including the above coefficient have been developed in
[11-13], extensions for gauge group SU(n) spontaneously broken to
maximal torus have been given in [15-17] (see also [14] where some
bosonic contributions to low-energy effective action have been found).

\section{N=4 super Yang-Mills theory in harmonic superspace}

As well known, the most powerful and adequate approach to investigate
the quantum aspects of supersymmetric field theories is formulation of
these theories in terms of unconstrained superfields carrying out a
representation of the supersymmetry. Unfortunately such a manifestly N=4
supersymmetric formulation for N=4 Yang-Mills theory is still unknown.
A purpose of this paper is study a structure of low-energy effective
action for N=4 SYM as a functional of N=2 superfield strengths. In this
case it is sufficient to realize the N=4 SYM theory as a theory of N=2
unconstrained superfields. It is naturally achieved within harmonic
superspace. The N=2 harmonic superspace [19] is the only manifestly N=2
supersymmetric formalism allowing to describe general N=2
supersymmetric field theories in terms of unconstrained N=2
superfields.  This approach has been successfully applied to problem of
effective action in various N=2 models in recent works [7, 9, 13, 16].

\noindent\hspace*{\parindent}From point of view of N=2 SUSY, the N=4
Yang-Mills theory describes interaction of N=2 vector multiplet with
hypermultiplet in adjoint representation. Within harmonic superspace
approach, the vector multiplet is realized by unconstrained analytic
gauge superfield $V^{++}$. As to hypermultiplet, it can be described
either by a real unconstrained superfield $\omega$
($\omega$-hypermultiplet) or by a complex unconstrained analytic
superfield $q^+$ and its conjugate ($q$-hypermultiplet). In the
$\omega$-hypermultiplet realization, the classical action of N=4 SYM
model has the form

$$
S[V^{++},\omega]=\frac{1}{2g^2}{\rm tr}\int d^4xd^4\theta
W^2- \frac{1}{2g^2}{\rm tr}\int
d\zeta^{(-4)}\nabla^{++}\omega \nabla^{++}\omega \eqno(1)
$$
The first terms here is pure N=2 SYM action and the second term is
action $\omega$-hypermultiplet. In $q$-hypermultiplet realization, the
action of the N=4 SYM model looks like this

$$
S[V^{++},q^+,\stackrel{\smile}{q}^+]= \frac{1}{2g^2}{\rm
tr}\int d^4xd^4\theta W^2- \frac{1}{2g^2}{\rm tr}\int
d\zeta^{(-4)}q^{+i}\nabla^{++}q^+_i \eqno(2)
$$
where
$$
q^+_i=(q^+,\stackrel{\smile}{q}^+),\qquad
q^{i+}=\varepsilon^{ij}q^+_j=(\stackrel{\smile}{q}^+,-q^+)
\eqno(3)
$$
All other denotions are given in [19].  Both models (1,2) are
classically equivalent and manifestly N=2 supersymmetric by
construction. However, as has been shown in [19], both these models
possess hidden N=2 supersymmetry and as a result they actually are N=4
supersymmetric.

\section{General form of non-holomorphic effective potential}

We study the effective action $\Gamma$ for N=4 SYM with gauge group
SU(2) spontaneously broken down to U(1). This effective action is
considered as a functional of N=2 superfield strengths $W$ and $\bar
W$. Then holomorphic effective potential ${\cal F}$ depends on chiral
superfield $W$ and it is integrated over chiral subspace of N=2
superspace with the measure $d^4x\,d^4\theta$. Non-holomorphic
effective potential ${\cal H}$ depends on both $W$ and $\bar W$. It is
integrated over full N=2 superspace with the measure $d^4x\,d^8\theta$.
Let us begin with dimensional analysis of low-energy effective action.
Taking into account the mass dimensions of $W$, ${\cal F}(W)$, ${\cal
H}(W,\bar W)$ and the superspace measures $d^4x\,d^4\theta$ and
$d^4x\,d^8\theta$ ones write

$$
{\cal F}(W)=W^2f\left(\frac{W}{\Lambda}\right),\qquad {\cal H}(W,\bar
W)={\cal H}\left(\frac{W}{\Lambda}, \frac{\bar W}{\Lambda}\right)
\eqno(4)
$$
where $\Lambda$ is some scale and $f(\frac{W}{\Lambda})$ and ${\cal
H}(\frac{W}{\Lambda},\frac{\bar W}{\Lambda})$ are the dimensionless
functions of their arguments.  Due to remarkable properties of N=4 SYM
in quantum domain, the effective action is scale independent. Therefore

$$
\Lambda\frac{d}{d\Lambda} \int
d^4x\,d^4\theta W^2f\left(\frac{W}{\Lambda}\right)=0, 
\quad
\Lambda\frac{d}{d\Lambda} \int d^4x\,d^8\theta{\cal
H}\left(\frac{W}{\Lambda}, \frac{\bar W}{\Lambda}\right)=0
\eqno(5)
$$
First of eqs (5) leads to $f(\frac{W}{\Lambda})=const$. Second of eqs
(5) reads

$$
\Lambda\frac{d}{d\Lambda}{\cal H}=g\left(\frac{W}{\Lambda}\right)+ \bar
g\left(\frac{\bar W}{\Lambda}\right)\eqno (6)
$$
Here $g$ is arbitrary chiral function of chiral superfield
$\frac{W}{\Lambda}$ and $\bar g$ is conjugate function.
The integrals of $g$ and $\bar{g}$ over full N=2 superspace vanish and
eqs(5) takes
place for any $g$ and $\bar g$.  Since $f(\frac{W}{\Lambda})=const$ the
holomorphic effective potential ${\cal F}(W)$ is
proportional to classical lagrangian $W^2$. General solution to eq (6)
is written as follows
$$
{\cal
H}\left(\frac{W}{\Lambda},\frac{\bar W}{\Lambda}\right)=
c\log\frac{W^2}{\Lambda^2}\log\frac{\bar W^2}{\Lambda^2}\eqno(7)
$$
with arbitrary coefficient $c$. As a result, holomorphic effective
potential is trivial in N=4 SYM theory. Therefore, namely
non-holomorphic effective potential is leading low-energy quantum
contribution to effective action.  Moreover, the non-holomorphic
effective potential is found exactly up to coefficient and given by eq
(7) [10]. Any perturbative or non-perturbative quantum corrections are
included into a single constant $c$.  However, this result immediately
face the problems:
\begin{itemize}
\item[1)] is there exist a calculational procedure
allowing to derive ${\cal H}(W/\Lambda,\bar W/\Lambda)$ in form (7)
within a model?
\item[2)] what is value of $c$? If $c=0$, the non-holomorphic
effective potential vanishes and low-energy effective action in N=4 SYM
is defined by the terms in effective action depending on the covariant
derivatives of $W$ and $\bar{W}$,
\item[3)] what is structure of
non-holomorphic effective potential for the other then SU(2) gauge
groups?
\end{itemize}

The answers all these questions have been given in [11-17]. Further we
are going to discuss a general manifestly N=2 supersymmetric and gauge
invariant procedure of deriving the non-holomorphic effective potential
in one-loop approximation [13,16]. This procedure is based on the
following points:
\begin{itemize}
\item[1)] formulation of N=4 SYM theory in terms of N=2
unconstrained superfields in harmonic superspace [19],
\item[2)] N=2
background field method [9] providing manifest gauge invariance on all
steps of calculations,
\item[3)] Identical transformation of path integral for
effective action over N=2 superfields to path integral over some N=1
superfields. This point is nothing more then replacement of variables
in path integral,
\item[4)] Superfield proper-time technique
(see first of refs [2]) which is
manifestly covariant method for evaluating effective action in
superfield theories.
\end{itemize}

Next section is devoted to some details of calculating non-holomorphic
effective potential.

\section{Calculation of non-holomorphic effective potential}

We study effective action for the classically equivalent theories (1,
2) within N=2 background field method [9]. We assume also that the
gauge group of these theories is SU(n). In accordance with background
field method [9], the one-loop effective action in both realizations of
N=4 SYM is given by

$$
\Gamma^{(1)}[V^{++}]=\frac{i}{2
}{\rm Tr}_{(2,2)} \log\stackrel{\frown}{\Box}-\frac{i}{2}{\rm
Tr}_{(4,0)} \log\stackrel{\frown}{\Box} \eqno(8)
$$
where
$\stackrel{\frown}{\Box}$ is the analytic d'Alambertian introduced in
[9].
$$
\begin{array}{rcl} \stackrel{\frown}{\Box}&=&{\cal
D}^m{\cal D}_m+ \displaystyle\frac{i}{2}({\cal D}^{+\alpha}W){\cal
D}^-_\alpha+ \displaystyle\frac{i}{2}(\bar{\cal D}^+_{\dot{\alpha}}\bar
W){\bar {\cal D}}^{-\dot{\alpha}}-\\ &-&\displaystyle\frac{i}{4}({\cal
D}^{+\alpha} {\cal D}^+_\alpha W){\cal
D}^{--}+ \displaystyle\frac{i}{8}[{\cal
D}^{+\alpha},{\cal D^-_\alpha}]W+\displaystyle\frac{i}{2}
\{\bar{W},W\} \end{array}\eqno(9)
$$
The formal definitions of the ${\rm
Tr}_{(2,2)}\log\stackrel{\frown}{\Box}$ and ${\rm
Tr}_{(4,0)}\log\stackrel{\frown}{\Box}$ are given in [13]. Our purpose
is finding of non-holomorphic effective potential ${\cal H}(W,\bar W)$
where the constant superfields $W$ and $\bar W$ belong to Cartan
subalgebra of the gauge group SU(n). Therefore, for calculation of
${\cal H}(W,\bar W)$ it is sufficient to consider on-shell background

$$
{\cal D}^{\alpha(i}{\cal D}^{j)}_\alpha W=0
\eqno(10)
$$
In this case the one-loop effective action (8) can be written in the
form [13]

$$
\exp(i\Gamma^{(1)})=\frac{\int{\cal D}{\cal
F}^{++}\exp\left\{ -\frac{i}{2}{\rm tr}\int d\zeta^{(-4)}{\cal
F}^{++}\stackrel{\frown} {\Box}{\cal F}^{(++)}\right\}} {\int {\cal
D}{\cal F}^{++}\exp\left\{-\frac{i}{2}{\rm tr}\int d\zeta^{(-4)}{\cal
F}^{++}{\cal F}^{++}\right\}}\eqno(11)
$$
The superfield ${\cal
F}^{++}$ belonging to the adjoint representation looks like ${\cal
F}^{++} = {\cal F}^{ij}u^+_iu^+_j$ with $u^+_i$ be the harmonics [19]
and ${\cal F}^{ij}={\cal F}^{ji}$ satisfy the constraints
$$
{\cal
D}^{(i}_\alpha{\cal F}^{jk)}=\bar{\cal D}^{(i}_{\dot{\alpha}} {\cal
F}^{jk)}=0,\qquad \bar{\cal F}^{ij}={\cal F}_{ij}\eqno(12)
$$

The next step is transformation of the path integral (11) to one over
unconstrained N=1 superfields. This point is treated as replacement of
variables in path integral (11). We introduce the N=1 projections of
$W$ ( see the details in [13, 16]). As a result one obtains

$$
\Gamma^{(1)}=\sum\limits_{k<l}\Gamma_{kl},\qquad \Gamma_{kl}=i{\rm
Tr}\log\Delta_{kl} \eqno(13)
$$
where
$$
\Delta_{kl}={\cal D}^m{\cal
D}_m-(W^{k\alpha}-W^{l\alpha}) {\cal
D}_\alpha+(\bar{W}^k_{\dot{\alpha}}-\bar{W}^l_{\dot{\alpha}}) \bar{\cal
D}^{\dot{\alpha}}+|\Phi^k-\Phi^l|^2 \eqno(14)
$$
and ${\cal D}_m$,
${\cal D}_\alpha$, $\bar{\cal D}_{\dot\alpha}$ are the supercovariant
derivatives. Here
$$
\Phi={\rm diag}(\Phi^1,\Phi^2,\dots,\Phi^n),\qquad
\sum\limits_{k=1}^n \Phi^k=0. \eqno(15)
$$
$$ W_\alpha={\rm
diag}(W_\alpha^1,\dots,W_\alpha^n),\qquad \sum\limits_{k=1}^n
W_\alpha^k = 0
$$
The operator (14) has been introduced in [16].  Thus, we get a problem
of effective action associated with N=1 operator (14). Such a problem
can be investigated within N=1 superfield proper-time technique.
Application of this technique leads to lowest contribution to effective
action in the form

$$
\Gamma_{kl}=\frac{1}{(4\pi)^2}\int d^8z\frac{W^{\alpha kl}W^{kl}_\alpha
\bar{W}^{kl}_{\dot{\alpha}}\bar{W}^{\dot{\alpha}kl}}
{(\Phi^{kl})^2(\bar{\Phi}^{kl})^2} \eqno(16)
$$
where
$$
\Phi^{kl}=\Phi^k-\Phi^l,\qquad W^{kl}=W^k-W^l \eqno(17)
$$
Eqs (13,
16, 17) define the non-holomorphic effective potential of N=4 SYM
theory in terms of N=1 projections of N=2 superfield strengths.  The
last step is restoration of N=2 form of effective action (16). For
this purpose we write contribution of non-holomorphic effective
potential to effective action in terms of covariantly constant N=1
projections $\Phi$ and $W_\alpha$
$$
\int d^4xd^8\theta{\cal
H}(W,\bar{W})=\int d^8zW^\alpha W_\alpha
\bar{W}_{\dot{\alpha}}\bar{W}^{\dot{\alpha}}\frac{\partial^4{\cal
H}(\bar{\Phi},\Phi)}{\partial\Phi^2\partial\bar{\Phi}^2}+ {\rm
derivatives} \eqno(18)
$$
Comparison of eqs (17) and (18) leads to
$$
\Gamma^{(1)}=\int d^4xd^8\theta{\cal H}(\bar{W},W)
$$
$$
{\cal H}(W,\bar{W})=\frac{1}{(8\pi)^2}\sum\limits_{k<l}
\log\left(\frac{(\bar{W}^k-\bar{W}^l)^2}{\Lambda^2}\right)
\log\left(\frac{(W^k-W^l)^2}{\Lambda^2}\right) \eqno(19)
$$
Eq (19) is our final result. In partial case of SU(2) group
spontaneously broken down to U(1) eq (19) coincides with eq (7) where
$c=1/(8\pi^2)$.

\section{Discussion}

Eq (19) defines the non-holomorphic effective potential depending on
N=2 superfield strengths for N=4 SU(n) super Yang-Mills theories. As a
result we answered all the questions formulated in section 3. First, we
have presented the calculational procedure allowing to find
non-holomorphic effective potential. Second, we calculated the
coefficient c in eq (7) for SU(2) group. It is equal to $1/(8\pi)^2$.
Third, a structure of non-holomorphic effective potential for the gauge
group SU(n) has been established.  It is interesting to point out that
the scale $\Lambda$ is absent when the non-holomorphic effective
potential (19) is written in terms of N=1 projections of $W$ and $\bar
W$ (see eqs (15, 16)). Therefore, the $\Lambda$ will be also absent if
we write the non-holomorphic effective potential through the components
fields. We need in $\Lambda$ only to present the final result in
manifestly N=2 supersymmetric form.  N=1 form of non-holomorphic
effective potential (16) allows very easy to get leading bosonic
component contribution. Schematically it has the form $F^4/|\phi|^4$,
where $F_{mn}$ is abelian strength constructed from vector component
and $\phi$ is a scalar component of N=2 vector multiplet. It means that
non-zero expectation value of scalar field $\phi$ plays a role of
effective infrared regulator in N=4 SYM theories.  Generalization of
low-energy effective action considered in [10-17] and containing all
powers of constant $F_{mn}$ has recently been constructed in [19] in
terms of N=2 superconformal invariants. The direct proof of absence of
three- and four-loop corrections to ${\cal H}$ was given in [20].

\noindent{\bf Acknowledgments.}
I am very grateful to E.I. Buchbinder, E.I. Ivanov, S.M. Kuzenko, B.A.
Ovrut, A.Yu Petrov, A.A. Tseytlin for collaboration and valuable
discussions. The work was supported in part by the RFBR grant
99-02-16617, RFBR-DFG grant 99-02-04022, INTAS grant 991-590, GRACENAS
grant 97-6.2-34, NATO collaborative research grant PST.CLG 974965 and
FAPESP grant. I thank Institute of Physics, University of Sao Paulo
for hospitality.

\end{document}